\documentclass[10pt]{iopart}
\usepackage[latin1]{inputenc}
\usepackage{graphicx}
\usepackage{iopams}
\usepackage{units}

\newcommand{\diffd}{\textrm{d}}

\begin{document}
\title{Single-particle-sensitive imaging of freely propagating ultracold atoms}
\author{R B\"{u}cker$^1$, A Perrin$^{1,2}$, S Manz$^1$, T Betz$^1$, Ch Koller$^1$, T~Plisson$^{1,3}$, J Rottmann$^4$, T Schumm$^{1,2}$, and J~Schmiedmayer$^1$}

\address{$^1$ Atominstitut, TU Wien, Stadionallee 2, 1020 Vienna, Austria}
\address{$^2$ Wolfgang Pauli Institute, Nordbergstrasse 15, 1090 Vienna, Austria}
\address{$^3$ Institut d'Optique, 2, avenue Augustin Fresnel, 91127 Palaiseau cedex, France}
\address{$^4$ Physikalisches Institut, Universit\"{a}t Heidelberg, Philosophenweg 12, 69120 Heidelberg, Germany}

\begin{abstract}
We present a novel imaging system for ultracold quantum gases in expansion. After release from a confining potential, atoms fall through a sheet of resonant excitation laser light and the emitted fluorescence photons are imaged onto an amplified CCD camera using a high numerical aperture optical system. The imaging system reaches an extraordinary dynamic range, not attainable with conventional absorption imaging. We demonstrate single-atom detection for dilute atomic clouds with high efficiency where at the same time dense Bose-Einstein condensates can be imaged without saturation or distortion. The spatial resolution can reach the sampling limit as given by the $\unit[8]{\mu m}$ pixel size in object space. Pulsed operation of the detector allows for slice images, a first step toward a 3D tomography of the measured object. The scheme can easily be implemented for any atomic species and all optical components are situated outside the vacuum system. As a first application we perform thermometry on rubidium Bose-Einstein condensates created on an atom chip. \textit{(published: New J. Phys. 11 (2009) 103039)}


\end{abstract}
\pacs{1315, 9440T}

\section{Introduction}

At the basis of all experimental research lies the ability to efficiently detect the phenomena one
wants to address. Consequently, the development of new detection techniques has often been the source of fundamental new
discoveries. For example, fifty years ago and thanks to single-photon detectors, Hanbury Brown and Twiss
(HBT) discovered photon bunching in light emitted by a chaotic source~\cite{HanburyBrown1956}, highlighting the importance of two-photon correlations and stimulating the development of modern quantum optics.

Within the field of ultracold atomic quantum gases, absorption imaging is the most popular detection technique. It provides quantitative access to the spatial distribution of atomic column densities in a single measurement. However, fundamental information regarding the atomic statistics and complex quantum correlations~\cite{Bloch2008} remains elusive, as absorption imaging becomes impracticable for low atomic densities (see \ref{sec:absorption}) and measuring column densities obscures correlations along the optical axis, reducing the signal-to-noise ratio significantly. 

In the last few years, considerable effort has been put into the development of single-atom detectors
to overcome such limitations. High-finesse cavities has been used to measure the quantum statistics of atom lasers~\cite{Oettl2005}, microchannel plate detectors have allowed the design of a position-sensitive three-dimensional single-atom detector for metastable He* atoms~\cite{Schellekens2005}. At the same time, fluorescence imaging schemes have been implemented to detect single atoms trapped in optical potentials~\cite{Schlosser2001,Kuhr2001,Nelson2007,Bakr2009}, in an atom beam~\cite{Bondo2006} or on atom chips \cite{Wilzbach2009} and more recently electron microscopy has been used to spatially resolve the position of individual atoms in an optical lattice~\cite{Gericke2008}.  However, apart from microchannel plates, which are only an option for atoms in highly excited metastable states, none of these techniques is able to image entire cloud samples in time-of-flight expansion.

In this article we report on a novel imaging scheme for quantum gases in expansion based on fluorescence photons emitted by atoms falling through a thin \emph{light sheet} \cite{Lett1988,Esslinger1996} which are projected onto an Electron Multiplying Charge Coupled Device (EMCCD) camera. This quasi-two-dimensional horizontal layer contains resonant probe light which is shined into the experimental setup symmetrically from two sides, all imaging elements are positioned outside the vacuum vessel. The detector is employed in an atom chip experiment \cite{Folman2002,Trinker2008a} mainly used for research on low-dimensional Bose gases \cite{Hofferberth2008} and matter-wave interferometry \cite{Schumm2005b}. 

Using the new detector we are able to image ultracold clouds with an extremely high dynamic range: single atoms forming a dilute cloud are distinctly visible, whereas at the same time dense Bose-Einstein condensates are imaged without any saturation or distortion. The spatial resolution depends on the excitation power used, for not too high intensities we are limited only by the pixel size of $\unit[8]{\mu m}$ in object space. The combination of spatial resolution and single-atom sensitivity allows for direct observation of correlation phenomena in cold gases as well as detection of samples or features which would be invisible to conventional imaging approaches. Furthermore, in addition to images integrated along the optical axis, the light sheet can be pulsed, realizing slice images, and hence enables in principle a 3D "tomography" of the object.

This article is organized as follows: in section~\ref{sec:concept} we discuss the concept of the detector, from the working principle to detailed experimental implementations and simulations of the imaging process. In section~\ref{sec:characterisation} we present various characterisation measurements, before proceeding to the visibility of single atoms in the acquired images in section~\ref{sec:sgl_at_detec}. Thermometry of Bose gases deep within the quantum degenerate regime is presented in section~\ref{sec:thermometry} as a first application which exploits the detector's extraordinary dynamic range. Finally, several more technical details as well as a discussion on the limits of absorption imaging for extremely dilute clouds are included in the appendices~A-C.

\section{Working principle of the imaging system}
\label{sec:concept}

\begin{figure}
	\centering\includegraphics[width=134mm]{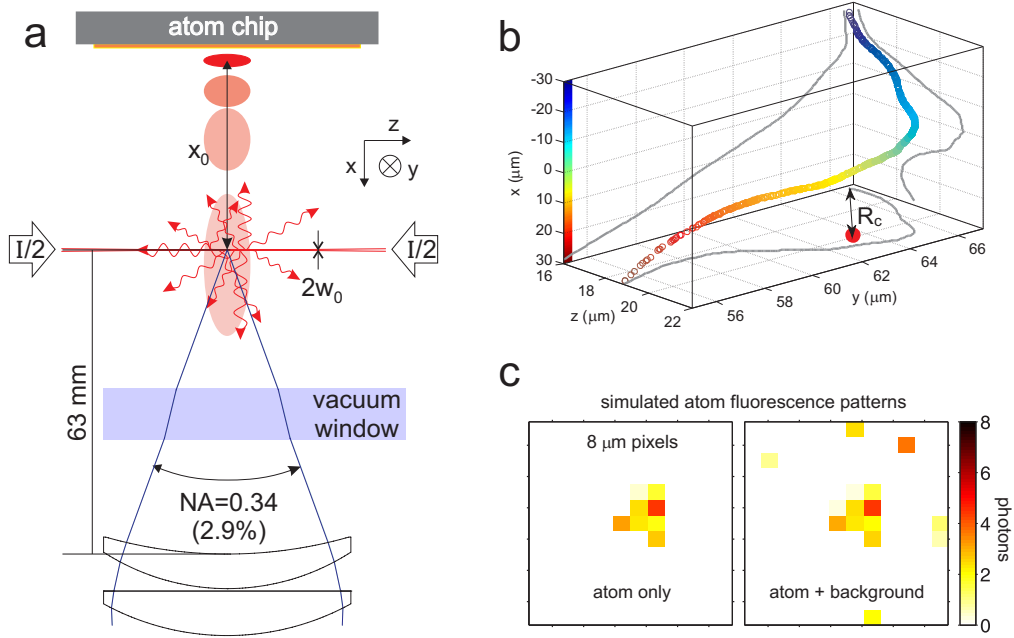}
	\caption{(a) Schematic of the light sheet system. The sheet (waist in x-direction $w_0=\unit[20]{\mu m}$) is formed by two counterpropagating laser beams, each beam carrying half the total intensity $I$. After falling a distance $x_0\sim \unit[10]{mm}$ (not drawn to scale) the expanding atom cloud pierces the sheet and emits fluorescence photons. Outside the vacuum, an objective with numerical aperture of $N=0.34$ collects 2.9\% of the light and transmits it to a detector. (b) Typical trajectory of a single atom falling through the sheet, obtained by Monte Carlo simulation. The atom performs a random walk in momentum space due to the stochastic photon absorption and emission events. This random walk translates into real space to a typical value of $\unit[10]{\mu m}$. The color encodes the vertical position of the atom. In the horizontal plane, the red dot indicates the transverse centroid position of the emitted photons, which is located at a distance $R_c$ from the initial atom position. (c) Simulated image of a typical single atom, including effects of diffusion, signal amplification and lens aberration (left) and background (right).}
	\label{fig:overview}
\end{figure}

In this section, we describe the principle of the new time-of-flight (TOF) fluorescence imaging. As a distinct feature, it applies the idea of using a thin horizontal light sheet as an excitation source~\cite{Lett1988,Esslinger1996} to single-atom sensitive, spatially resolved detection of ultracold quantum gases.
As an atom cloud released from the atom chip trapping potentials falls through the sheet (see figure \ref{fig:overview}a), a fraction of the spontaneously emitted fluorescence photons is collected by imaging optics and finally detected on an amplified EMCCD camera. In contrast to absorption imaging (or other schemes that rely on imaging of the initial light beam itself), where the background level is given by photon shot noise (see \ref{sec:absorption}), for fluorescence images it is solely given by technical noise sources and stray light. Thus, an arbitrarily high signal-to-background ratio can be reached if the technical background level is low compared to the photon number detected from individual atoms.

In section \ref{sec:principle} we will concentrate on the experimental implementation and some technical details necessary to achieve this goal, whereas in section \ref{sec:design} various properties of the system are discussed and analysed using Monte Carlo simulations of the imaging process.

\subsection{Experimental implementation}
\label{sec:principle}

All experiments are performed in an atom chip setup for the creation and manipulation of $^{87}$Rb Bose-Einstein condensates~\cite{Wildermuth2004,Schumm2005b,Trinker2008a,Hofferberth2008}. At the end of each experimental cycle, the atomic cloud is released from the chip trap by rapidly extinguishing the magnetic fields which provide an anisotropic harmonic confinement with trapping frequencies of typically 20~Hz longitudinally and 2~kHz transversally. Falling under gravity, the cloud expands according to its momentum distribution, which is anisotropic for the cigar-shaped Bose-Einstein condensates produced in our experiment, but isotropic in the case of a thermal gas \cite{Ketterle1998}. The cloud usually contains on the order of $10^4$ atoms and its temperature ranges from 20~nK to $2~\mu$K.

After $\unit[40-50]{ms}$ time-of-flight expansion, corresponding to about 10~mm falling distance, the atoms pass a quasi-two-dimensional light sheet, tuned resonant or near-resonant to the $^{87}$Rb D$_2$ line $F=2\rightarrow F'=3$ cycling transition at $\unit[780]{nm}$.\footnote{In case the initial cloud is in the lower $F=1$ hyperfine ground state, a small amount of light tuned to the $F=1\rightarrow F'=2$ transition has to be added to the light sheet, in order to pump atoms into the bright $F=2$ state. The dependence of the total signal on the pumping light power first rises steeply until it saturates at a value typically much smaller than the imaging light power.} To cancel the net radiation pressure arising from a single resonant beam, the sheet consists of two identical, but counterpropagating beams, created by two identical optics assemblies outside the experiment chamber. Each of these consists of a fibre outcoupler, creating a collimated beam of waist $d_0=\unit[4.5]{mm}$, followed by a cylindrical singlet lens which shapes a highly anisotropic elliptical Gaussian beam with a vertical waist of $w_0=\unit[20]{\mu m}$. The usable area of the light sheet is limited by the beam waist in horizontal direction of $d_0=\unit[4.5]{mm}$ and by the Rayleigh length with respect to the vertical waist of $z_0=\unit[1.6]{mm}$, resulting in a sufficiently homogeneous detection area with a diameter of $\unit[3]{mm}$. A lin$\perp$lin polarization scheme is used in order to avoid a standing wave configuration.  It takes about $100~\mu$s for a single atom to cross the light sheet and on the order of 1~ms for a typical Bose-Einstein condensate.

Positioning of the beams is accomplished by observing the fluorescence of magnetically trapped rubidium atoms when illuminated with the light sheet from several directions. Once one of the beams is placed correctly with respect to the position of the falling atoms, the other one can be precisely adjusted by optimizing the re-coupling of each of the sheet beams into the opposite fibre. A re-coupling efficiency of up to 60\% has been achieved, reducing the expected amount of background signal, as the re-coupled light is guided away from the experiment.

The fluorescence light emitted from the atoms is collected by a high-numerical-aperture (NA) lens underneath the vacuum chamber. The objective is custom-built, following design principles as in \cite{Alt2002}, but re-designed to provide a higher NA of $N=0.34$ (i.e., 2.9\% solid angle coverage) at a working distance of 60~mm. These values are close to the constraints given by the experiment geometry, which limits the opening angle to $N\sim 0.35$. The depth of field is just large enough to encompass the light sheet without compromising the system's resolution. On the other hand, the high NA effectively blurs any spurious signal originating from outside the light sheet (see section \ref{sec:background}). The fluorescence light is then refocused onto a $512\times512$ pixel EMCCD camera\footnote{Andor iXon+ DU-897} with a twofold magnification. The field of view is about $\unit[4\times 4]{mm^2}$ at a pixel size of $\unit[8\times 8]{\mu m^2}$ in object space. While the objective is designed to be near-diffraction-limited on the optical axis, off-axis aberrations (in particular field curvature) affect the performance towards the edges. Still, the optical resolution is sampling-limited over the entire detection area of $\sim\unit[3]{mm}$ diameter, as given by the light sheet parameters (see section \ref{sec:resolution} for a more detailed discussion on the effective resolution properties).

The back-illuminated, frame-transfer EMCCD detector placed in the image plane provides 72\% quantum efficiency at a wavelength of $\unit[780]{nm}$. In an EMCCD camera an \emph{electron multiplication register}, placed in the signal chain between pixel array and readout unit, amplifies the signal. It consists of a chain of pixel potential wells, through which the photoelectrons are transported sequentially, as through the readout register of a conventional CCD. However, to accomplish the transport, a much higher voltage on the order of 50~V is applied, increasing the probability of impact ionization in the destination well to about 1\%. Having passed all of the $\sim500$ wells, the initial charge has been amplified by a factor of up to 1000, ensuring the visibility of even single photons over the readout noise. As a drawback, signal statistics are complicated due to the stochastic amplification process, effectively doubling the signal variance \cite{Basden2003}. In \ref{app:calib} we lay out the calibration procedure of the camera used to reconstruct photon number expectation values from the amplifier output. The remaning major technical noise caused by the camera are clock-induced charges (CIC) which are created in the pixel row shifting during readout (see section \ref{sec:background} for a quantitative discussion).

After laying out the construction basics of the light sheet system we will now turn to identifying suitable operation parameters for achieving single-atom detection sensitivity as well as high spatial resolution.

\subsection{Design parameters and simulations}
\label{sec:design}

Both the signal strength and the spatial resolution of the system depend on the geometrical and optical parameters of the light sheet. While this is obvious for the signal strength, the resolution's dependence arises from the random scattering force fluctuations exerted on the atoms within the light sheet. These impose a random walk in momentum space, which propagates into real space on a timescale given by the interaction time of a single atom with the sheet, as illustrated in figure~\ref{fig:overview}b. As a consequence, each \emph{atom fluorescence pattern}, i.e. the image contribution due to a single atom, has an irregular, blurred shape. Furthermore, a single atom pattern has a random spatial offset $R_c$ with respect to the undisturbed atom position (\emph{centroid deviation}) due to early scattering events as indicated in figure~\ref{fig:overview}b. 

The requirements of strong signal and low spatial diffusion are somewhat contrary as the first calls for a high number of scattered photons, whereas for the latter a low number of random walk steps (or even some cooling molasses effect) is desirable, thus some compromise has to be found. This reflects in the hardware-implemented light sheet vertical waist of $w_0=\unit[20]{\mu m}$ which was designed to be as thin as possible while retaining sufficient signal and detection area (i.e., Rayleigh length). It allows the minimization of the propagation time within the sheet, thus reducing visible diffusion, and moreover to keep the required depth of field low. The sheet position is variable, but large times of flight on the order of $t_{\rm{TOF}}\approx\unit[45]{ms}$ are used to zoom into the momentum distribution and further reduce interaction time due to the higher vertical atom velocity. Sensible values for the optical parameters are a light sheet power corresponding to a peak intensity on the order of the atomic transition's saturation $I_S=\unit[3.58]{mW/cm^2}$ (for random polarization, see section~\ref{sec:signal_size}), while a slight red detuning (within the linewidth of the atomic transition) leads to Doppler cooling, reducing spatial diffusion in the light sheet direction while still maintaining sufficient signal (see section \ref{sec:acf-psf}).

We employ a Monte Carlo simulation of individual atoms in the light sheet to analyse the aforementioned atom fluorescence patterns. Their individually different shapes are due to the involved stochastic processes (diffusion, photon shot noise, EM amplification), a typical example is shown in figure \ref{fig:overview}c. For typical light sheet parameters, we expect each atom to scatter about 800 photons (see section \ref{sec:signal_size}), about 15 of which contribute to each fluorescence pattern, considering solid angle coverage and quantum efficiency of the detector. The simulated photons are spread over typically 7 detector pixels (see section \ref{sec:single_atom_signal}). For the given values, our simulation estimates a deviation of the centroid position of the fluorescence pattern with respect to the undisturbed atom positions on the order of $R_c\approx\unit[3]{\mu m}$ RMS. The typical radius of the patterns is on the order of $R_a\approx\unit[6]{\mu m}$ RMS, including imaging and detector contributions. This results in a total effective resolution limit of below $\unit[10]{\mu m}$ (all values given in object space coordinates). In the next section these estimations will be compared to calibration measurements.

\section{Characterisation}
\label{sec:characterisation}

Having explained the principle and construction of the light sheet fluorescence imaging, we now turn to the results of several characterisation measurements performed with the system. Data on background and signal strength and an estimation on the achievable spatial resolution will be presented.

\subsection{Background}
\label{sec:background}

\begin{figure}%
\centering\includegraphics{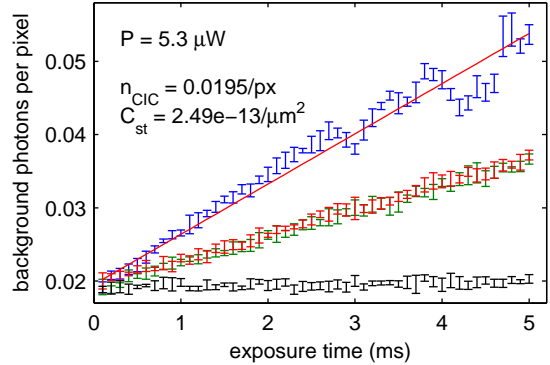}%
\caption{Characterisation of background level. The plot shows the mean photon number per pixel with both sheets activated (blue), each single sheet (red, green), and no light (black), as a function of exposure time. A linear fit to the slope corresponding to both sheets results in a stray light constant of $C_{st}=\unit[2.49\cdot10^{-13}]{\mu m^{-2}}$ as defined in the text. The independence of the no-light curve on exposure time at a level of $n_{\rm{CIC}}=\unit[0.0195]{/pixel}$ indicates the absence of any other stray light source than the light sheet itself.}%
\label{fig:background}%
\end{figure}

A crucial quantity concerning the sensitivity of the system is the mean amount of background photons\footnote{In the interest of readability we will refer to any detector output as (effective) "photons", whether the respective contribution is due to actual photons hitting the CCD or to any of the different camera noise sources that are described in this subsection. See \ref{app:calib} for a discussion of the signal chain and calibration of the EMCCD camera.} per pixel. It can be described by $\sigma_b=\sqrt{2\cdot n_b(P,t_e,n_a)}$, where the factor of two originates from the amplification process in the electron multiplying register \cite{Basden2003}. The background photon number $n_b$ can be expressed by a heuristic formula:

\begin{equation}
n_b(P,t_e,N_a)=n_{\rm{CIC}} + C_{st} A_{px} \eta_d t_e P/\hbar \omega + A_{px} \eta_r \eta_d / \pi (2x_0 N)^2 \cdot n_a N_a.
\label{eq:backphot}
\end{equation}

Each term in this equation corresponds to one of the three main contributions to the background signal, which typically are on the same order of magnitude:

\begin{itemize}
	\item As mentioned before, there is a certain unavoidable amount of camera noise due to \emph{clock-induced charges} (CIC). In contrast to thermally excited electrons (dark current), which can be suppressed by sufficient cooling of the detector, CIC photoelectrons do not depend on exposure time, as they are excited by impact ionization during readout \cite{Robbins2003}. They obey a Poissonian distribution with a mean of $n_{\rm{CIC}}$.
	\item Even though all optical elements in the beam path of the light sheet have anti-reflection coatings, we expect a certain amount of stray light scattered onto the CCD. This signal should depend linearly on the power in the light sheet $P$, exposure time $t_e$, detection efficiency $\eta_d$ as given by the numerical aperture and quantum efficiency, and object-space pixel size $A_{px}$. Its magnitude per pixel is expressed in the factor $C_{st}$ of dimension of inverse area. $\hbar\omega$ denotes the energy per photon. 
	\item Finally, at about 1~cm distance from the light sheet the Au-coated atom chip acts as a mirror with reflectivity $\eta_r\approx0.95$ for the light emitted by the atoms. The reflected images of the atoms are far out of focus, thus, we expect a homogeneous background contribution from scattered and reflected light, which is proportional to the total signal in the image, given by the product of total atom number $N_a$, scattered photons per atom $n_a$, and detection efficiency $\eta_d$. Furthermore it is expected to scale reciprocally with the size of the blur spot due to defocus as given by the distance of the focal plane from the chip $x_0$ and the numerical aperture $N$. This geometrical estimation complies within 15\% to experimental results obtained by comparing background images to the outside parts of images containing atoms.
\end{itemize}

It is important to note, that we expect all those contributions to be approximately homogeneous over the field of view, as any stray light sources as well as the mirrored atoms lie far out of the focus of the system. This yields a blur spot, over which the background photons are distributed, which is much larger than the detection area. 

The first two contributions can be measured directly from background images without atoms. Results for $n_b$ are shown in figure \ref{fig:background} and indicate that the background signal is well below 0.1 photons per pixel. However, the additional noise due to the stochastic amplification process of the EMCCD detector has to be taken into account when relating background level to its standard deviation, i.e. a typical background of $n_b=0.05$ photoelectrons mean leads to a noise of about $\sigma_b=(2\cdot0.05)^{1/2}\approx0.3$ photons per pixel. 

Whereas this is significantly less than the single-atom signal of about $n_a\sim 15$ photons at typical settings, some care has to be taken when comparing those numbers, as $n_a$ refers to the photons contained in a full fluorescence pattern, which is spread over $\sim7$~pixels for usual settings (see figure \ref{fig:overview}c and section \ref{sec:single_atom_signal}). The fluorescence patterns typically have a peaked distribution over those pixels, assuming an equal distribution instead yields a lower bound to an effective single-atom signal-to-background ratio. It relates the typically 15 photons originating from a single atom to the standard deviation of the noise signal integrated over 7 pixels, which is usually less than one photon:

\[
{\rm SBR}_{at}\gtrsim \frac{n_a}{\sqrt{2\cdot7\cdot n_b}} \approx 18.
\]

\subsection{Signal strength}
\label{sec:signal_size}

The amount of photons emitted by each atom while falling through the light sheet is expected to depend linearly on the total power in the light sheet up to the point where saturation occurs. However, in contrast to an atom at rest, the saturation effect does not completely suppress a further increase of the signal for an atom falling through the light sheet. For high powers the scattering rate will still rise towards the edges of the Gaussian intensity profile of the light sheet and the effective interaction time will increase further. The slope of the signal-versus-power curve will only gradually decrease for powers beyond saturation. Assuming a two-level atom interacting with a resonant light sheet, the total signal per atom can be written as

\begin{equation}
n_a(P)=\eta_d \bar{v}^{-1} \frac{\gamma}{2} \int_{-\infty}^{\infty} \frac{P/P_S\cdot \rme^{-2x^2/w_0^2}}{1+P/P_S\cdot \rme^{-2x^2/w_0^2}} \diffd x,
\label{eq:photperat}
\end{equation}

where $P$ denotes the power in the two beams together, $\gamma$ the inverse lifetime of the atomic excited state and $\bar{v}$ the mean velocity of the atoms while passing the sheet. The saturation power $P_S$ is given by the saturation intensity of the used transition and the area of the light sheet beam. For our geometry, it can be calculated as $P_S = \pi d_0 w_0 / 2 \cdot I_S \approx \unit[5]{\mu W}$ with the horizontal waist of the light sheet $d_0=\unit[4.5]{mm}$ and a saturation intensity of $I_S=\unit[3.58]{mW/cm^2}$ \cite{Steck2001}. It has been assumed that the atom moves quickly enough through the polarization gradient of the lin~$\bot$~lin-polarized sheet that the polarization can be seen as isotropic.

\begin{figure}
	\centering
		\includegraphics[width=85mm]{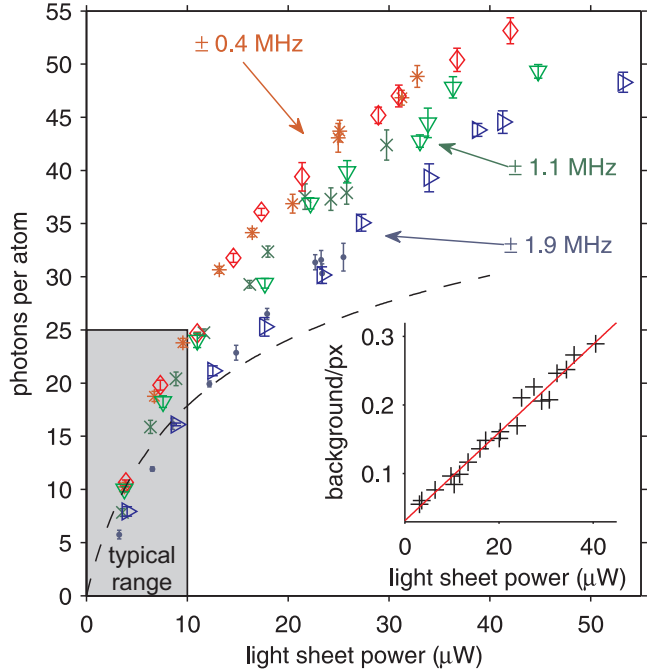}
	\caption{Detected photons per atom as a function of light sheet power for different detunings. The total atom number was $\sim10^4$, calibrated by absorption imaging. Note that the power range shown in the figure exceeds typical values ($P \lesssim \unit[10]{\mu W}$) in order to emphasize the saturation effect. As expected, the plots for each pair of positive/negative detuning come to lie on top of each other. However, the absolute photon numbers as well as the saturation power exceed the values expected from the naive model presented in the text, shown as black dashed line for resonant light. The inset shows the background signal per pixel within the range of the measurement in the main figure (5~ms exposure time, see section \ref{sec:background}).}
	\label{fig:total_signal}
\end{figure}

To measure the actual dependence of the per-atom signal on light sheet power in the experiment, a reference for the total atom number within a cloud is necessary. As the atom number fluctuations between each experimental run are below 10\%, this is readily available by switching back and forth between light sheet and conventional absorption imaging, the latter providing an absolute mean atom number of clouds created under specific experimental conditions. Results are shown in figure \ref{fig:total_signal}. Whereas the measured values roughly correspond to the theoretical estimation for typical powers ($P \lesssim \unit[10]{\mu W}$), the signal is significantly higher than expected for powers far beyond saturation. We attribute this to an insufficiency of the naive model outlined above, which neglects the multi-level structure of the atom as well as any coherent processes. Furthermore, slight mechanical misalignments or optical imperfections (diffraction artifacts, non-ideal Gaussian mode) of the light sheet might lead to an increased detection volume and thus contribute to the deviation.

\subsection{Resolution}
\label{sec:resolution}

\begin{figure}
	\centering
		\includegraphics[width=120mm]{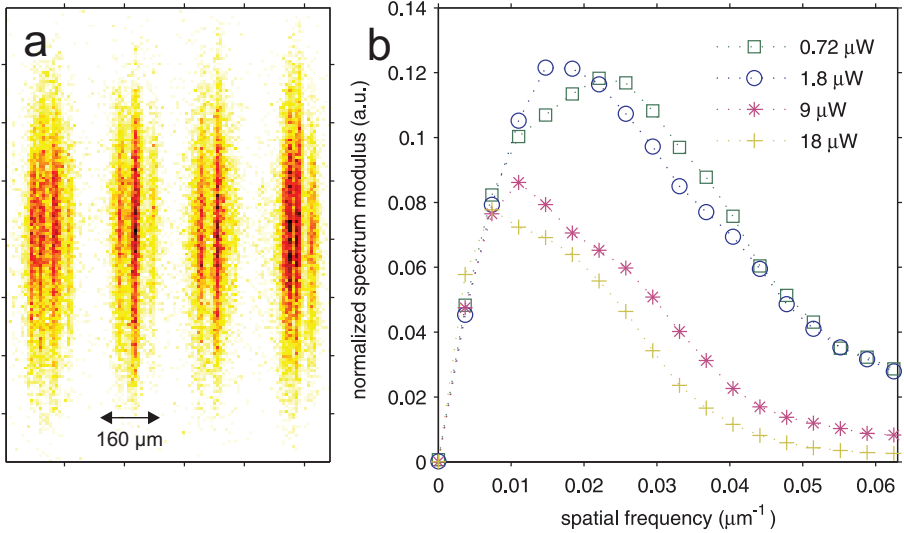}
	\caption{(a) Four typical images of degenerate atomic clouds exhibiting density modulations after time-of-flight expansion ($\unit[1.8]{\mu W}$ light sheet power). The spectrum of these density modulations as recorded by the imaging system is used to estimate its spatial frequency response. (b) Spectra (FFT modulus) of images as shown in (a) for different light sheet powers. Each series corresponds to an average over 100 single image spectra. Before calculating the spectra, the respective mean profile of each dataset has been subtracted to remove the contribution of the cloud shape envelope. Furthermore, each spectrum is normalized to the mean total signal in its dataset. The non-vanishing value at the Nyquist limit of $\nu_N=\unit[0.0625]{\mu m^{-1}}$ indicates, that an increase in resolution could be obtained by using optics with higher magnification.}
	\label{fig:ripples_resolution}
\end{figure}

To characterize the imaging system, it is crucial to be able to estimate its spatial resolution. This task is nontrivial in our case, because there are two distinct contributions: (1)  The imaging optics, its resolution, depth of focus and the sampled detection in the CCD \cite{Hadar1997} and (2) a significant contribution by the diffusive motion of the atoms in the light sheet itself which causes the emitted light to come from an off centred extended region, as laid out in section~\ref{sec:design}. Thus only an \emph{in situ} method is appropriate, which uses data acquired in the experiment itself. Furthermore, as the image position of each individual atom may deviate from the initial atom position (atom image centroid deviation $R_c$ as defined in section~\ref{sec:design}), a \emph{collective} structure within the imaged quantum gas has to be chosen as a resolution target. Such a structure is readily available in our experiment: the atom chip traps used are typically of a highly anisotropic cigar shape with aspect ratios on the order of 100. In such geometries, even at temperatures well below the condensation temperature $T_C$ of a homogeneous gas, phase fluctuations remain present along the longitudinal direction of the cloud~\cite{Dettmer2001}. These fluctuations propagate into density modulations in time-of-flight expansion, which are clearly visible in the images of clouds (figure~\ref{fig:ripples_resolution}a). The spectrum of spatial frequencies contained in the modulations extends well beyond the Nyquist limit as given by the pixel size \cite{Imambekov2009a}, so they can be used as a broadband target to estimate the system's resolution limit. 

The measured spatial spectrum along the modulation direction can be described as a product of the initial spectrum and an unknown modulation transfer function (MTF), which specifies the system's frequency response, and thus its resolution. Averaged spectra, obtained from 100 images each, are shown in figure~\ref{fig:ripples_resolution}b for different light sheet powers. They are corrected for the contribution of the cloud shape envelope and normalized to the total signal in the respective images. 

To deduce the complete MTF, a full theoretical description of the initial modulation spectra for our experimental parameters would be necessary, which is not yet available. However, as the initial physical system is unchanged between the different datasets, different light sheet settings can directly be compared to each other, and we obtain the \emph{relative} values of the respective transfer functions. Atom diffusion inside the light sheet (see figure \ref{fig:overview}b) is clearly observed in terms of a power-dependent magnitude decrease of the measured spectra. Furthermore, as an important result it can be noted that for low powers even at the Nyquist limit of $\nu_N=\unit[0.0625]{\mu m^{-1}}$ the measured spectrum does not vanish, i.e. features spaced by twice the pixel size $2d_{px}=\unit[16]{\mu m}$ are still separable. This means that in this regime (i.e. for low excitation) the resolution is sampling limited, and could be enhanced further by increasing the optical magnification, thus increasing $\nu_N$.

\section{Single atom detection}
\label{sec:sgl_at_detec}

In this section we address the question of the visibility of single atoms in the fluorescence images. We first focus on proving the sensitivity of the detector to single atoms and then present a method to quantitatively describe the single-atom signal properties.

\subsection{Single atom autocorrelation}
\label{sec:acf-psf}

To assess the visibility of single atoms in an image of an entire cloud, it is necessary to decompose the image data into components with distinct typical length scales. Exactly this is conveniently accomplished by investigating the two-dimensional autocorrelation function defined as (see also \cite{Altman2004}): 

\[
a(\Delta x,\Delta y)=\sum_x\sum_y S(x,y) S(x+\Delta x,y+\Delta y),
\]

where $S(x,y)$ denotes the background-corrected image and the summation limits are defined by the subset of pixels allowing for a pairwise spacing of $(\Delta x, \Delta y)$. This function is zero outside the region defined by the spatial extent of the imaged cloud. Inside this region, structures of different typical extents in the initial image become visible as contributions to the autocorrelation with individual correlation lengths.

\begin{figure}
\centering\includegraphics[width=134mm]{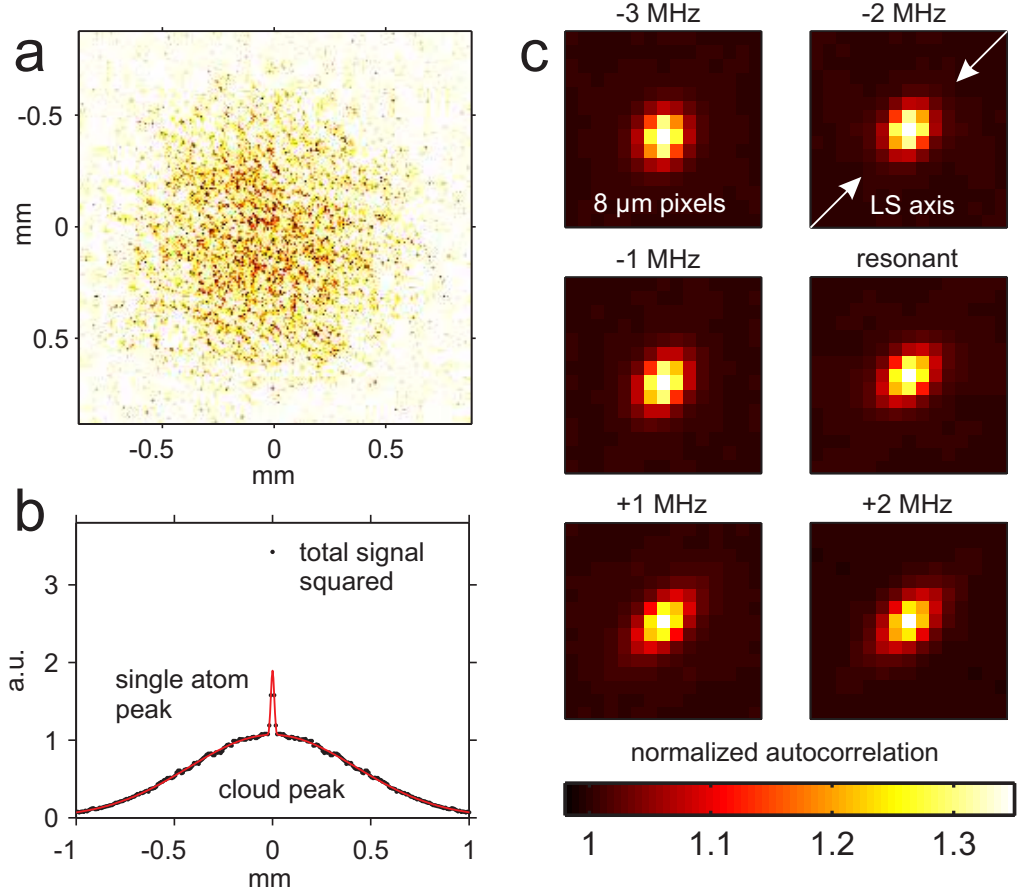}
\caption{Image autocorrelation and single-atom molasses effect. (a) Typical image of a thermal cloud (fully integrated along the vertical direction) using the light sheet system. (b) Cut through the y-axis of the two-dimensional autocorrelation function of (a). Three features can be distinguished which correspond to photon shot noise, single atoms and the cloud envelope. A double Gaussian fit is shown as red line. (c) Central peak of the normalized two-dimensional autocorrelation function for thermal cloud images at different detuning settings of the light sheet. The scattering rate per atom was held constant by adjusting the intensity in the sheet. The anisotropy of the peak depends on the detuning, indicating a one-dimensional Doppler cooling during the imaging process. Arrows indicate the direction of the light sheet (LS).}
\label{fig:autocorr}
\end{figure}

For a light sheet image of a thermal cloud as shown in figure \ref{fig:autocorr}a, three such contributions become visible in the autocorrelation function (figure \ref{fig:autocorr}b shows a cross section). Firstly, a high, delta-shaped peak at $(\Delta x,\Delta y)=(0,0)$ corresponds to the squared total signal in the image, including correlated and uncorrelated (photon shot noise) structures. Secondly, the structure size given by the entire cloud accounts for a broad peak with Gaussian shape. Finally, a peak with short but finite structure size emerges on top of the cloud peak if the system is well focused (see \ref{sec:focus}). We attribute this peak to the visibility of the single-atom structure within the cloud. Its width on the order of $\unit[10]{\mu m}$ corresponds well to the Monte Carlo simulation results as shown in figure \ref{fig:overview}c and can be considered as a mean atom fluorescence pattern. Note, though, that the shape of this peak only gives very limited information about the effective spatial resolution of the system, as it contains no contribution by the centroid deviation $R_c$ (see section ~\ref{sec:design}). Furthermore, the height of the central pixel is obstructed by the uncorrelated signal contribution. Two-atom correlations (e.g. as in~\cite{Schellekens2005}) can in principle be detected, but are much weaker in amplitude than the single atom peak and will be discussed in an upcoming article.

In the next step of data treatment we isolate the average single-atom pattern by using an entire series of images made under identical conditions. We divide the mean autocorrelation function by the autocorrelation function of the averaged image. In the averaged image, the single-atom structure vanishes, but other contributions to the autocorrelation remain. We obtain the \emph{normalized} autocorrelation function, which only contains contributions by the single-atom structure and is equal to unity elsewhere.

In figure \ref{fig:autocorr}c it is shown how the shape of the mean fluorescence pattern depends on the detuning from resonance of the light sheet. While changing the light frequency, the scattering rate of the atoms was being held constant by adjusting the beam intensity. An anisotropy of the fluorescence pattern is observed and can readily be explained by the enhanced absorption of photons along the light sheet axis (which is under $45^\circ$ angle with respect to the pixel grid) which exerts a stochastic force as large as the one by emission, which in contrast is isotropic. As expected, the anisotropy almost vanishes when using light which is red-detuned by up to half the natural linewidth. This fact shows the observation of a Doppler molasses effect on the single-atom level.

\subsection{Single atom signal}
\label{sec:single_atom_signal}

\begin{figure}
	\centering\includegraphics[width=\linewidth]{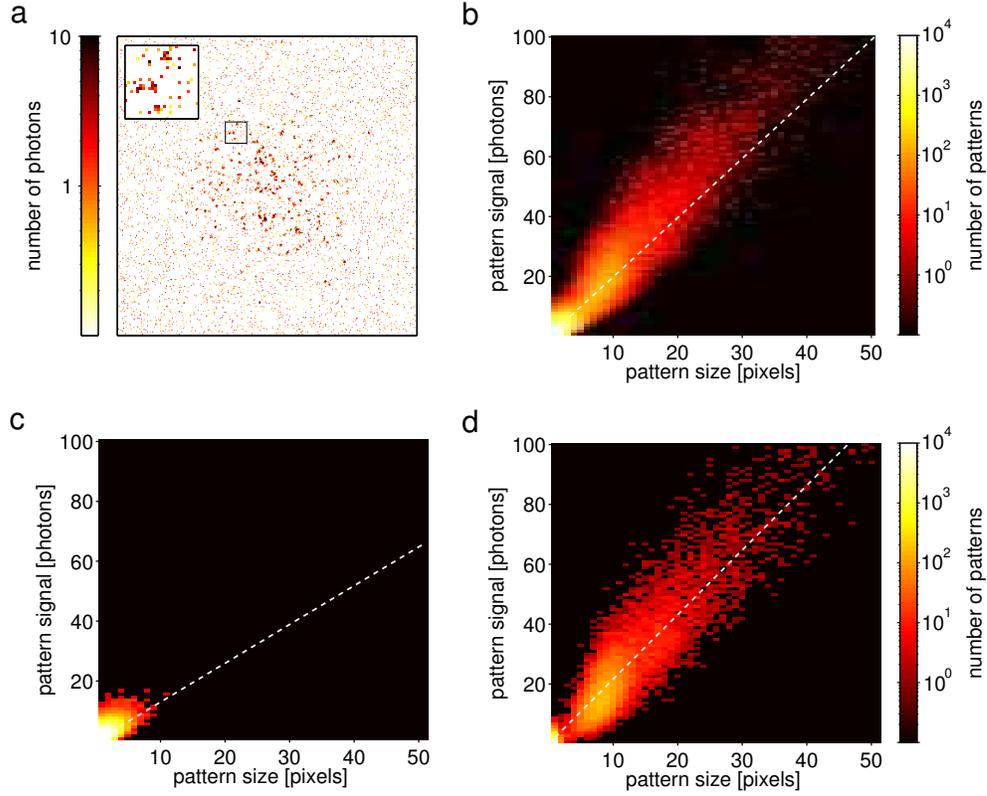}
	\caption{(a) Example of a simulated light sheet image containing the fluorescence scattering of 200 atoms and obtained with light sheet beams of 8~$\mu$W and detuning of $-1$~MHz from the atomic transition. The simulations include the effect of the background noise. (b) Distribution $\mathcal{P}\left(s_p,n_p\right)$ of the size $s_p$ and signal $n_p$ of fluorescence patterns obtained from the treatment of 100 simulated images. The white dashed line is a plot of the linear equation $n_p=\alpha s_p$, where $\alpha\simeq1.98$~photons per pixel is the mean number of photons per pixel in atomic scattering patterns (see text). (c) Distribution $\mathcal{P}$ obtained from 100 background images. In this case, $\alpha\simeq1.30$. (d) Same as (b), but obtained from images with a clean background. In this case, $\alpha\simeq2.16$.}
	\label{fig:sim_dist_2d}
\end{figure}

\begin{figure}
	\centering\includegraphics[width=\linewidth]{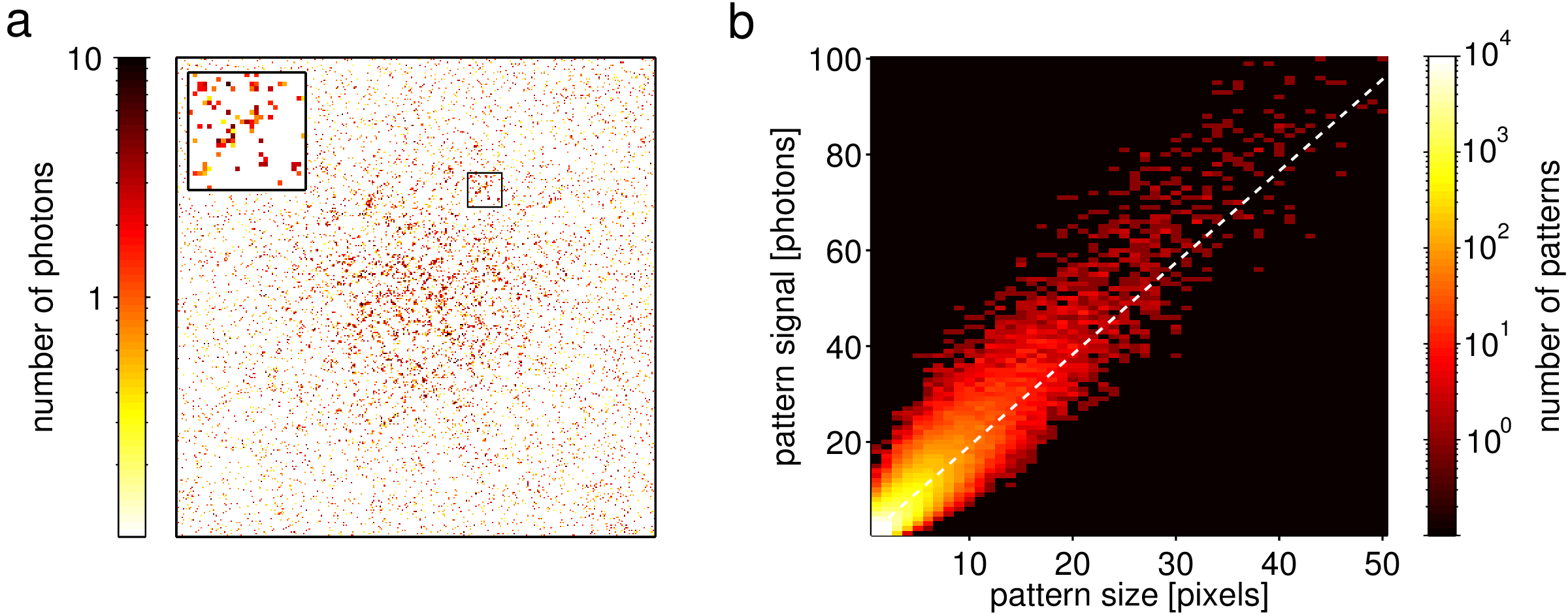}
	\caption{(a) Typical experimental light sheet image containing about 200 atoms at $1~\mu$K. The light sheet parameters are the same as for figure~\ref{fig:sim_dist_2d}a. (b) Distribution $\mathcal{P}\left(s_p,n_p\right)$ of the size $s_p$ and signal $n_p$ of fluorescence patterns obtained from the treatment of 100 experimental images. The white dashed line is a plot of the linear equation $n_p=\alpha s_p$, where $\alpha\simeq1.91$~photons per pixel is the mean number of photons per pixel in atom fluorescence patterns.}
	\label{fig:exp_dist_2d}
\end{figure}

Having demonstrated the sensitivity of the imaging system to individual atoms we will now characterize single-atom signal properties. For a very dilute atomic cloud, the fluorescence signals of single atoms appear as distinct localized patterns on an image (see figure~\ref{fig:overview}c). Looking at the statistics of these patterns, more precisely at their size $s_p$ and at the total number of photons they contain $n_p$, it is in principle possible to characterize more accurately the interaction of the atoms with the light sheet than just by analysing quantities given by the total signal in an image (see section \ref{sec:signal_size}). In order to explore the possibilities and limitations of such an approach we have developed an algorithm able to locate each pattern within an image. Using this algorithm we are able to deduce the joint distribution $\mathcal{P}\left(s_p,n_p\right)$ of size of and total signal within the patterns out of a given set of images. Figure~\ref{fig:sim_dist_2d}b shows a typical distribution for a set of 100 images obtained by Monte Carlo simulations. Each image contains the fluorescence signal of 200 single atoms at a temperature of $0.8~\mu$K. For such parameters, it is very unlikely that two atom fluorescence patterns overlap (see figure~\ref{fig:sim_dist_2d}a) and one can identify each pattern with the fluorescence scattering of a single atom if the background noise level in each image is negligible. This is generally not the case since stray light photons as well as clock-induced charges (see section \ref{sec:background}) can also produce patterns spread over several pixels on the image. Fortunately, the shape of the distribution due to this background signal is generally very different from the one due to the atoms (see figure~\ref{fig:sim_dist_2d}c and figure~\ref{fig:sim_dist_2d}d). This allows the selection of those patterns which are most likely due to atoms. It can be used in turn to filter the experimental images and remove the background noise to some extent. In the case of the simulated images of figure~\ref{fig:sim_dist_2d}, we are able to retrieve more than 90\% of the atomic scattering patterns with a minimum confidence of 95\%~\footnote{For each pair $(s_p,n_p)$ the confidence level is defined as one minus the ratio between the number of patterns with this value pair for background and signal images.}. For the set of experimental images analysed in figure~\ref{fig:exp_dist_2d}, the same treatment allows us to retrieve about 75\% of the atomic scattering patterns with a minimum confidence of 95\%. We attribute the difference between these two results to the naive model used to realize the Monte Carlo simulations, but also to instabilities of the experimental setup which may cause a slight further bluring of the atomic fluorescence patterns.

For sufficiently low background level the shape of the distribution $\mathcal{P}$ is mostly governed by the fluorescence scattering of single atoms, and some interesting physical quantities can be retrieved from its analysis. For example, the average photon density of the atomic fluorescence patterns, $\alpha$, can be obtained by analysing the ratio $n_p/s_p$ of the signal and size of the different observed patterns. More precisely, one can define $\alpha$ as the least square solution of a linear regression performed on the properties ($n_p$,$s_p$) of each pattern of the studied set of images. We find a value of 1.98~photons per pixel in the case of the simulated images after a careful filtering of the background patterns. This value is close to the one obtained for the experimental images, 1.91~photons per pixel, and differs only slightly from the one we obtain from the analysis of simulated images with a clean background, 2.16~photons per pixel (see figure~\ref{fig:sim_dist_2d} and \ref{fig:exp_dist_2d}).


Another very interesting quantity that could in principle be deduced from the analysis of the distribution $\mathcal{P}$ is the mean number of photons per atom. Such a quantity would provide a complementary method to the calibration presented in section \ref{sec:signal_size}, which would not rely on comparison with a separate imaging system. In the case of the simulated images as shown in figure \ref{fig:sim_dist_2d}, one can roughly estimate a value of 18 photons for the mean number of photons per atom from the position of the peak of the distribution $\mathcal{P}$. This value is close to the expected one of 17.9 that we directly extract from the simulations. This result could probably be improved by developing more accurate analytic models of the scattering of the atoms that we could use to fit the distribution $\mathcal{P}$. For experimental images, this method fails because of the absence of a distinct peak in the observed distribution $\mathcal{P}$. Again, inaccuracy of the simple two-level model used for the simulations and experimental instabilities might explain the discrepancy between simulations and experiments.

\section{Thermometry of quantum gases}
\label{sec:thermometry}

One of the advantages of the light sheet fluorescence detector over conventional methods lies in its extraordinarily high dynamic range. Extremely low as well as relatively high atomic densities can be measured simultaneously with the same accuracy. An obvious example of such a case is a condensed cloud well below the critical temperature, where the dense condensate is surrounded by a thermal fraction which only consists of a few atoms. As even such dilute parts remain visible, fitting the wings of the density distribution remains possible in a temperature and thermal fraction range that extends well beyond what is feasible using absorption imaging \cite{Gerbier2004a}. Furthermore, for typical temperatures the signal of the thermal fraction is strong enough to perform a temperature fit on single shots (even tomographic shots) instead of averaged data. This makes light sheet imaging a versatile tool for thermometry at low temperatures, which is especially critical for the one-dimensional regimes studied in our experiment~\cite{Hofferberth2007b,Petrov2000,Mazets2008}.

\begin{figure}[tbp]
	\centering\includegraphics[width=134mm]{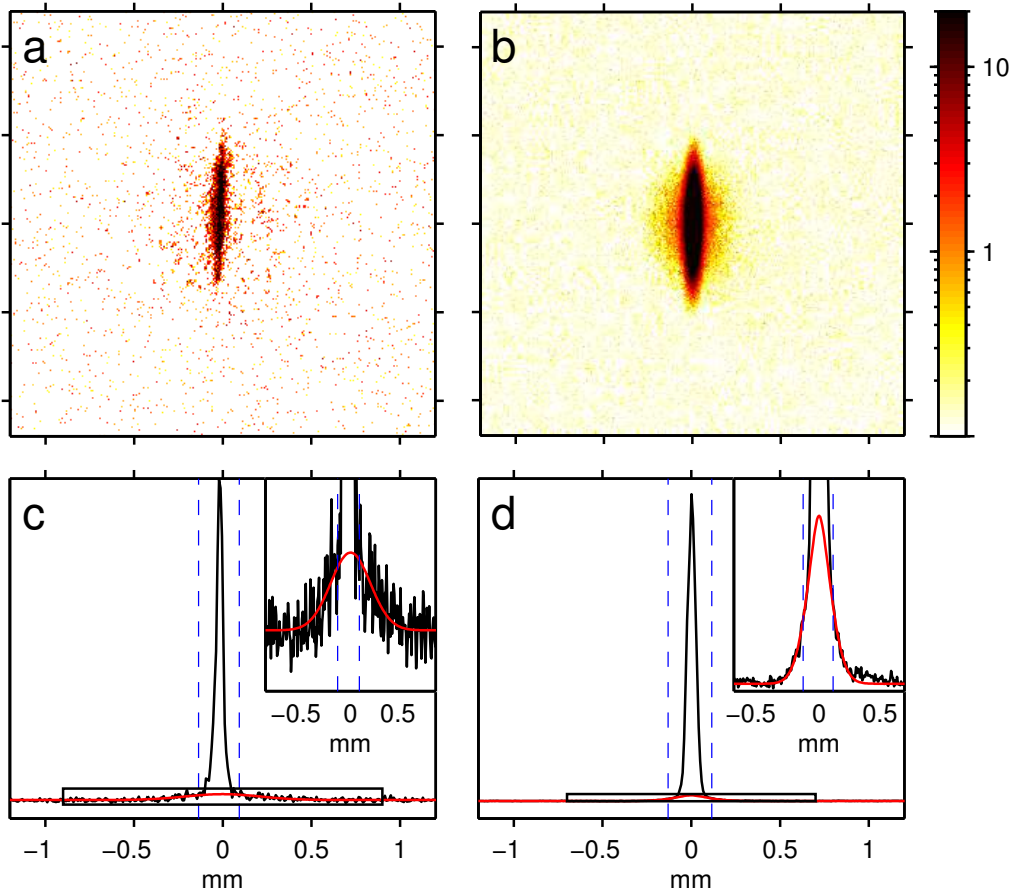}
	\caption{(a) Single-shot image of a $400~\mu$s time-slice at the centre of a quasi-pure Bose-Einstein condensate. (b) Averaged image of a quasi-pure Bose-Einstein condensate over 20 realisations. (c) Longitudinal profile of (a). Even though the thermal fraction is very small, its shape can still be fitted in a single experiment run due to the high dynamic range of the fluorescence detector. We deduce a temperature of 220~(150)~nK and estimate the thermal fraction to include the fluorescence scattering of about 100 atoms. The inset shows a zoom into the shape of the thermal fraction. (d) Longitudinal profile of (b). By fitting the thermal fraction we are able to obtain a temperature of only 64~(15)~nK. In figures (c) and (d) the vertical dashed lines indicate the boundary of the excluded region of the fit.}
	\label{fig:long_prof}
\end{figure}

In figure~\ref{fig:long_prof}a, a fit to the shape of the thermal fraction surrounding a $400~\mu$s time-slice of a quasi-pure Bose-Einstein condensate in a single shot is shown. Even though the thermal fraction is composed of only about 100 atoms, we are still able to deduce a temperature of 220~(150)~nK. This corresponds to a ratio $T_c/T$ on the order of 6, where $T_c$ is the critical temperature of the gas. We also estimate the condensed fraction to represent more than 90~\% of the atoms~\footnote{It is interesting to note that for such trap parameters and temperature the maximum number of thermal atoms of a non-interacting Bose gas is 150, which agrees fairly well with what we observe.}. Figure~\ref{fig:long_prof}b similarly shows a fit to data averaged over 20 images of atoms from a more elongated trap at an even lower temperature of 64~(15)~nK. In this case $T_c/T$ is also about 6 and the condensed fraction represents $93~\%$ of the atoms.

\section{Summary and outlook}

We have implemented and characterized a novel imaging scheme based on atom fluorescence in a light sheet and demonstrated its many advantages for time-of-flight imaging compared to conventional schemes based on absorption. We have in particular demonstrated its high dynamic range which combines sensitivity to single atoms with the ability to image dense Bose-Einstein condensates. 

A careful study of the stochastic scattering process an atom undergoes when travelling through the light sheet and a detailed analysis of all background sources yields a lower bound for the single-atom signal-to-background ratio of better than 18. We have additionally implemented an algorithm combining knowledge on the number and spatial distribution of fluorescence photons originating from single atoms to filter experimental images and retrieve $>70\%$ of individual atoms with $95\%$ confidence.

Phase fluctuations in elongated Bose-Einstein condensates created on atom chips transform into density fluctuations in time-of-flight expansion which have been used as an in situ broadband target to characterize the system resolution \cite{Imambekov2009a}. Using low excitation powers, the spatial resolution was limited only by the pixel size of 8~$\mu$m in object space. Higher powers were shown to reduce resolution due to atom diffusion, whereas at the same time the signal-to-background ratio is increased. Thus, excitation power, as well as detuning which was shown to cause a Doppler molasses effect, can be used to tailor the detection to a given system under study.

As a first application example we have efficiently performed thermometry of ultracold Bose gases deep in the quantum degenerate regime. We have also pointed out the possibility to image only time-slices of the atomic column-density. The latter adds effectively three-dimensional resolution to the light sheet fluorescence detector. 

We are convinced that this new scheme will open up many new research possibilities and quickly establish itself as a universal tool in cold atom research. It can be easily adapted for any atomic species; all optical components are situated outside the vacuum vessel. Specifically, it can readily be implemented in atom chip setups where the presence of the chip usually complicates standard imaging approaches.
The extraordinary bandwidth of the detector, including spatially resolved single-atom detection, paves the way to new ways of probing degenerate quantum gases such as the study of density-density correlations in time-of-flight, an increasingly active field of research~\cite{Schellekens2005,Jeltes2007,Perrin2007,Greiner2005,Hellweg2003,Rom2006,Oettl2005}. The study of low-dimensional physics in single sample realizations necessitates the detection of low atomic densities~\cite{Petrov2000}. Finally, the imaging process itself may provide insight into basic questions of atom-light interaction, especially by observing laser cooling effects on the single-atom level. 

\ack
We thank Alex Gottlieb for stimulating discussions, J\"{u}rgen Appel for his help in developing the recognition algorithm, and David A. Smith for critical reading of the manuscript. RB acknowledges support from the Studienstiftung des deutschen Volkes and the ESF EuroQUASAR program, AP from the Seventh Framework Programme under grant agreement n°236702, SM and RB from the FWF project CoQuS No. W1210-N16, TB from the SCALA network and CK from the FunMat research alliance.

\appendix

\section{Limitations of absorption imaging in time-of-flight}
\label{sec:absorption}

In our experiment, imaging of very dilute gases after long times of flight is desirable, as the spatial distribution of the atoms then fully reflects the momentum distribution in the initial cloud and is spread out wide enough to observe this quantity with good resolution. More specifically, measurements of correlation functions even necessitate the visibility of single atoms sparsely distributed in the falling cloud. In this appendix we will motivate the need for the imaging system as described in this paper to tackle such tasks, by estimating the limits of conventional low-power absorption imaging.\footnote{In the following, we will not regard other imaging schemes, such as phase contrast imaging~\cite{Andrews1997a} or saturated absorption imaging~\cite{Reinaudi2007}, as they are tailored to dense, trapped clouds. For both techniques, the ratio between signal photon number and incident intensity (which determines the shot noise) is lower than for low-power absorption imaging, which makes them less sensitive for dilute, expanding gases.} We will first turn to resolution limitations in time-of-flight before analysing the detection limit in the case of sparse atoms.

\subsection{Finite depth of field}
\label{sec:dof}

A general issue in imaging is the finite longitudinal extent of the volume in which objects can be imaged sharply. Any coherent or incoherent system suffers from quality degradation, once object and image distance do not meet the lens law $s_o^{-1}+s_i^{-1}=f_e^{-1}$, where $s_o,s_i$ and $f_e$ denote the object distance, image distance and focal effective length, respectively. This of course also affects the imaging of a translucent object with finite depth along the optical axis, like a cold atom cloud in time-of-flight. For the case of coherent absorption imaging, we derive an expression for the effective modulation transfer function (MTF) in \ref{app:MTF}, which -- in Fourier space -- links the column density $\rho_c(x,y)$ to the detected signal: 

\begin{equation}
M_{abs}(\nu_t)=\Theta\left(N \lambda^{-1} - \nu_t\right) \cdot e^{-\frac{\pi^2}{2} \lambda^2 R_z^2 \nu_t^4} \cdot  \textrm{cos} (\pi \lambda \Delta \nu_t^2),
\label{eq:absMTF}
\end{equation}

where we use the symbols $\nu_t=\sqrt{\nu_x^2+\nu_y^2}$ as the transverse spatial frequency in the object plane, $N$ as the numerical aperture, $\lambda$ as the wavelength, $R_z$ as the longitudinal RMS radius of the imaged cloud and $\Delta$ as the defocus with respect to the cloud center. $\Theta$ denotes the Heaviside step function. The interesting term in this equation is the $R_z$-dependent cutoff at $\nu_{max}\sim 1/\sqrt{\pi \lambda R_z}$, which means that at a cloud depth larger than $R_z\sim\lambda/\pi N^2$, the system is not limited by diffraction anymore, but by finite depth effects. The effective blur disk radius is then roughly given by $R_b\approx 1.22\cdot(2 \nu_{max})^{-1}$. Furthermore, stopping down the system to a lower numerical aperture does not enhance the resolution for thick objects, as opposed to an incoherent system.

For a longitudinal cloud extension in time-of-flight of $R_z \approx \unit[55]{\mu m}$ (which is reached after $t_{\rm TOF}\approx\unit[10]{ms}$ in a typical chip trap with radial trap frequency on the order of $2\pi\cdot\unit[1]{kHz}$), we get $\nu_{max}\approx\unit[0.086]{\mu m^{-1}}$ and $R_b\approx \unit[7.1]{\mu m}$, which is equivalent to the diffraction limit of a system with numerical aperture 0.067, i.e. a larger aperture will not significantly improve the image resolution. 

\subsection{Sensitivity}

Neglecting any technical source of noise (readout, optical impurities), the sensitivity of absorption imaging is limited by the photon shot noise of the imaging light irradiated onto the detector. If we assume unity detection efficiency and an effective detection area $A_{\rm{PSF}}$ given by the total point spread function corresponding to the resolution of the system, we can express the photon shot noise as 

\[
\sigma_b=\left(I_R/\hbar\omega\cdot t_e \cdot A_{\rm PSF}\right)^{1/2},
\]
where $I_R$ denotes the imaging light intensity at frequency $\omega/2\pi$ and $t_e$ the exposure time. On the other hand, the number of photons absorbed from the beam per atom is given by

\[
n_a=I_R/\hbar\omega\ \cdot t_e \cdot \sigma_0,
\]
with the cross section $\sigma_0$ of the used atomic transition. From those relations, it becomes clear, that it is desirable to use a high intensity and exposure time, while reducing the detection area as much as possible, i.e. increasing the system's resolution. Unfortunately, there are limits for the values of all those quantities. While for in situ imaging, $A_{\rm PSF}$ is given by the numerical aperture of the used objective and can in this case be approximated by $A_{\rm PSF}\approx\pi(1.22\lambda/2N)^2$, in time-of-flight the resolution is predominantly limited by the finite depth of field, as pointed out in the previous section. For a typical TOF cloud extension of $R_z=\unit[55]{\mu m}$, we get $A_{\rm PSF} \approx \unit[160]{\mu m^2}$. 

The integrated photon flux $I_R/\hbar\omega \cdot t_e$ cannot be increased arbitrarily either. The initial intensity must be kept well below saturation $I_S=\frac{\hbar\omega\Gamma}{2\sigma_0}$ (with the transition linewidth $\Gamma$) to retain a linear relation between atom number and total signal. On the other hand, the exposure time has to be short compared to any movement of the atoms due to gravity or light forces that occur during the imaging process, as this will cause blurring or Doppler shift for movement perpendicular to or along the imaging axis, respectively. 

A straightforward calculation shows, that the ratio between $n_a$ and $\sigma_b$ will not exceed unity even for intensities on the order of saturation (which will hinder easy quantitative image evaluation) and exposure times limited by gravity. To give a quantitative example, $^{87}$Rb atoms released from a typical chip trap after $\unit[10]{ms}$ time-of-flight, and a cloud depth as above, will result in an upper limit of $n_a=445$ photons per atom at $I_R=I_S/2$ and $t_e=\unit[35]{\mu s}$, whereas the corresponding shot-noise limited background is $\sigma_b=450$ photons. Interestingly, even narrowing the detection depth (e.g. by using spatially varying pumping schemes as in \cite{Andrews1997}) does not significantly increase this ratio, as the decrease in $A_{\rm{PSF}}$ will also lead to a shorter maximum value of $t_e$, which is necessary to keep the gain in resolution. Therefore, we cannot expect to be able to detect single atoms or even very dilute clouds using absorption imaging in time-of-flight, even more so as the perfect conditions as modelled here can hardly be met in the actual experiment and even higher times of flight would be desirable.\footnote{We have chosen the rather short expansion time of $\unit[10]{ms}$ to yield an upper bound for absorption imaging performance. At longer times as comparable to light sheet imaging ($\unit[40-50]{ms}$), resolution and sensitivity become significantly worse as $R_z$ increases.} Thus, any scheme to image ultracold clouds in time-of-flight with single-atom sensitivity has to rely on fluorescence imaging within a sufficiently thin detection depth, e.g. using a light sheet as demonstrated in this paper.

\subsection{Derivation of the effective MTF for absorption imaging}
\label{app:MTF}

For a cloud having a Gaussian density distribution with RMS radius $R_z$ along the optical axis and a constant transverse profile, the density at any point within the cloud is given by 

\[
\rho(x,y,z)=\frac{\rho_c(x,y)}{\sqrt{2\pi}R_z} \cdot \rme^{-\frac{z^2}{2R_z^2}},
\]

where $\rho_c(x,y)$ denotes the column density in transverse direction. As we are assuming coherent imaging light, the signal linearly transmitted by the optical system is \emph{not} the intensity, but the complex amplitude of a field. We thus have to define an object function $o(x,y,z)$, quantifying the amplitude transmittance density at a position $(x,y,z)$ within the imaging beam, which is modeled as a plane wave. For the object function, we do not take into account the attenuation of the spatial zero frequency Fourier component, i.e. the initial imaging light (weak object approximation). Furthermore, we neglect any interaction of the atoms with light already diffracted by preceding atoms and assume unity energy in the beam. We can then write the object function as

\begin{equation}
o(x,y,z)=\left[1/V -C \cdot \frac{\rho_c(x,y)}{\sqrt{2\pi}R_z} \cdot \rme^{-\frac{(z-\Delta)^2}{2R_z^2}}\right] \rme^{-\rmi kz},
\label{eq:object}
\end{equation}

where $\Delta$, $V$ and $k=2\pi\lambda^{-1}$ are the defocus of the trap centre with respect to the object plane, the mode volume of the imaging beam, and the wave vector of the imaging light, respectively. $C$ is a positive, real, dimensionless constant expressing the strength of the light scattering by the atoms.

In Fourier space, this translates to

\begin{equation}
\fl O(\nu_x,\nu_y,\nu_z)=1/V \cdot \delta(\nu_x,\nu_y,\nu_z+\lambda^{-1}) - C \rho'_c(\nu_x,\nu_y) \rme^{-2 \pi^2 R_z^2 (\nu_z+\lambda^{-1})^2} \rme^{i 2 \pi (\nu_z+\lambda^{-1}) \Delta},
\label{eq:object_spec}
\end{equation}

where $\rho'_c(\nu_x,\nu_y)$ denotes the spatial spectrum of the column density. The three-dimensional coherent modulation transfer function for a circular aperture is in paraxial approximation given by \cite{Gu2000}

\begin{equation}
H(\nu_t,\nu_z)=N^2/\lambda \cdot \Theta\left(N \lambda^{-1} - \nu_t\right) \delta\left(\nu_z-\lambda/2 \cdot \nu_t^2+\lambda^{-1}\right),
\label{eq:CMTF}
\end{equation}

where $N$ is the numerical aperture of the system and $\nu_t=\sqrt{\nu_x^2+\nu_y^2}$. The Heaviside step function, which expresses the transverse diffraction limit, is denoted by $\Theta(\nu)$. To yield the signal on the detector (we assume unity magnification) we have to calculate the six-dimensional Fourier integral over the squared modulus of $[H\cdot O](\nu_x,\nu_y,\nu_z)$ at $z=0$. If we assume the nontrivial term in equation \ref{eq:object_spec} to be small compared to unity, consistent with our initial weak object approximation, we can neglect interference terms between nonzero Fourier components and gain a linear relation between the transferred column density and the detected real space signal $S$:

\begin{eqnarray}
S(x,y) \propto [1-C' \int \Theta\left(N \lambda^{-1} - \sqrt{\nu_x^2+\nu_y^2}\right) \rho'_c(\nu_x,\nu_y) e^{-\frac{\pi^2}{2} \lambda^2 R_z^2 (\nu_x^2+\nu_y^2)^2} \nonumber\\
\times \textrm{cos} (\pi \lambda \Delta (\nu_x^2+\nu_y^2)) e^{2\pi i (\nu_x x+\nu_y y)} \diffd \nu_x \diffd \nu_y].
\label{eq:final_intensity}
\end{eqnarray}

In the limit of $\lambda\rightarrow 0$, corresponding to neglecting all wave optics effects, this result reduces to the classical ("shadow image") Lambert-Beer law expanded to first order, if we identify the coefficient $C' \propto C$ with the scattering cross section $\sigma_0$:

\[
S(x,y) \propto [1-\sigma_0 \rho_c(x,y)].
\]

From equation \ref{eq:final_intensity} it becomes clear that the effective transfer function between $\rho'_c$ and the detected atom signal spectrum is given by:

\begin{equation*}
M_{abs}(\nu_t)=\Theta\left(N \lambda^{-1} - \nu_t\right) \cdot e^{-\frac{\pi^2}{2} \lambda^2 R_z^2 \nu_t^4} \cdot  \textrm{cos} (\pi \lambda \Delta \nu_t^2).
\end{equation*}

The three factors in this function can readily be identified with the diffraction limit, finite depth falloff and coherent defocusing behaviour, respectively. Whereas the last factor reproduces the Talbot effect with self-imaging at $\Delta_T=2(\lambda\nu^2)^{-1}$, the second one is responsible for the resolution reduction of expanding clouds in time-of-flight as described in section \ref{sec:dof}.

\section{Focusing}
\label{sec:focus}
\begin{figure}%
\centering\includegraphics[width=80mm]{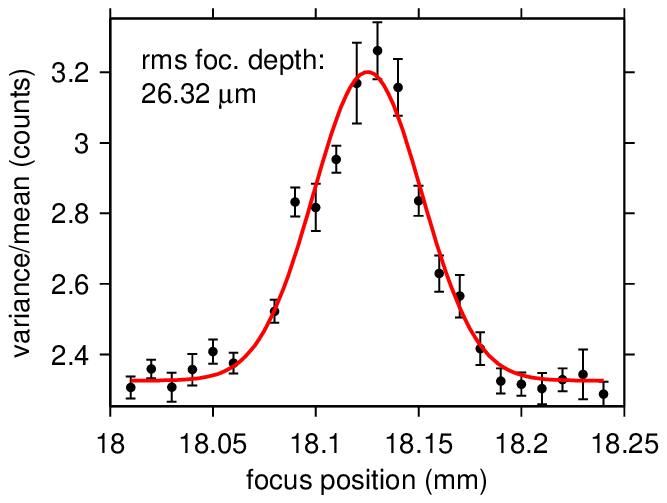}%
\caption{Focusing of the imaging system. The function shown is the normalized variance of a section within a thermal cloud (similar to the one shown in figure \ref{fig:autocorr}a) with approximately homogeneous density, as a function of the focus position. Out of focus, the value is close to 2, which corresponds to pure photon and amplification noise, plus a residual inhomogeneity. It rises considerably in vicinity of the focused position, as now the granularity due to the single atoms becomes visible. A Gaussian fit reveals an effective depth of focus of $\unit[26]{\mu m}$.}%
\label{fig:focus}%
\end{figure}

Focusing the light sheet system is accomplished by moving the entire detector up and down, using a micrometer-resolution translation stage. After a coarse adjustment, a very sensitive focus calibration can be made by taking images of small (on the order of 1000 atoms) thermal clouds. For a completely defocused system, the variance of the resulting image in the approximately homogeneous central part of the cloud is solely given by the photon shot noise of the emitted fluorescence light, the amplification noise and a residual inhomogeneity. A value for the variance divided by the mean of slightly above 2 is expected. Once the object-side focal plane of the system approaches the light sheet centre, this value begins to rise, as there is now also a contribution from the single-atom grain structure, as shown in figure \ref{fig:focus}. If set on the maximum value, the system is focused with an accuracy better than $\unit[5]{\mu m}$, as compared to the depth of field on the order of $\unit[10]{\mu m}$.

\section{EMCCD calibration}
\label{app:calib}

In this appendix we lay out the procedure to reconstruct expectation values for the number of photons impinged on each pixel from the amplified output of the CCD detector. Following the signal chain, the recorded fluorescence and stray light photons get converted to \emph{primary} photoelectrons in the CCD pixel array. Already during the transport of those to the amplification stage, the CIC electrons mentioned in section \ref{sec:background} are added and are thus inherently indistinguishable from actual photons. The amplification process as briefly described in section \ref{sec:concept} amplifies the primary electrons by a -- usually large -- number of \emph{secondary} electrons. As this process is a stochastic one, it has to be described by a probability distribution function of secondary electron numbers for a given count of primary electrons. Finally, the secondary electrons are converted to a voltage and digitized by a readout unit (RU) which itself adds a certain amount of readout noise. 

The signal $S_0$ finally obtained in a given pixel of an image acquired with the EMCCD camera is a measure of the initial number of electrons received by the RU. In our case it is expressed in terms of {\it counts}, an arbitrary unit introduced by the RU during the analog to digital conversion. With our settings, each count corresponds to 10.63~secondary electrons put out by the electron multiplier. This last value sets the binning of the output signal into counts. $S_0$ can generally be divided into three different contributions:

\begin{itemize}
\item The number of secondary electrons $n_s$ produced in the gain register of the camera from an initial number $n_p$ of photo-electrons. The relation between $n_s$ and $n_p$ is purely stochastic, as in any electron multiplying device. For a sufficiently high EM gain $g$, the probability distribution for $n_s$ can be written~\cite{Basden2003}
\begin{equation}\label{eq:em_stat}
\mathcal{D}\left(n_s;n_p,g\right)={\rm P}\left(n_s=x|n_p,g\right)=\frac{x^{n_p-1}\exp\left(-x/g\right)}{g^{n_p}(n_p-1)!}.
\end{equation}
It is worth noting that a good estimate of $n_p$ knowing $n_s$ is given by $n_s/g$~\cite{Basden2003}.
\item The readout noise $\sigma_{ro}$. It is a random noise with a zero mean value and a standard deviation of about 2~counts for our usual settings.
\item The reference baseline $b$. This contribution is just an offset set by the RU in order to avoid negative values of $S_0$. It can be considered constant for each pixel of a given image but may fluctuate from image to image. 
\end{itemize}

\begin{figure}[tbp]
	\centering\includegraphics[height=8cm]{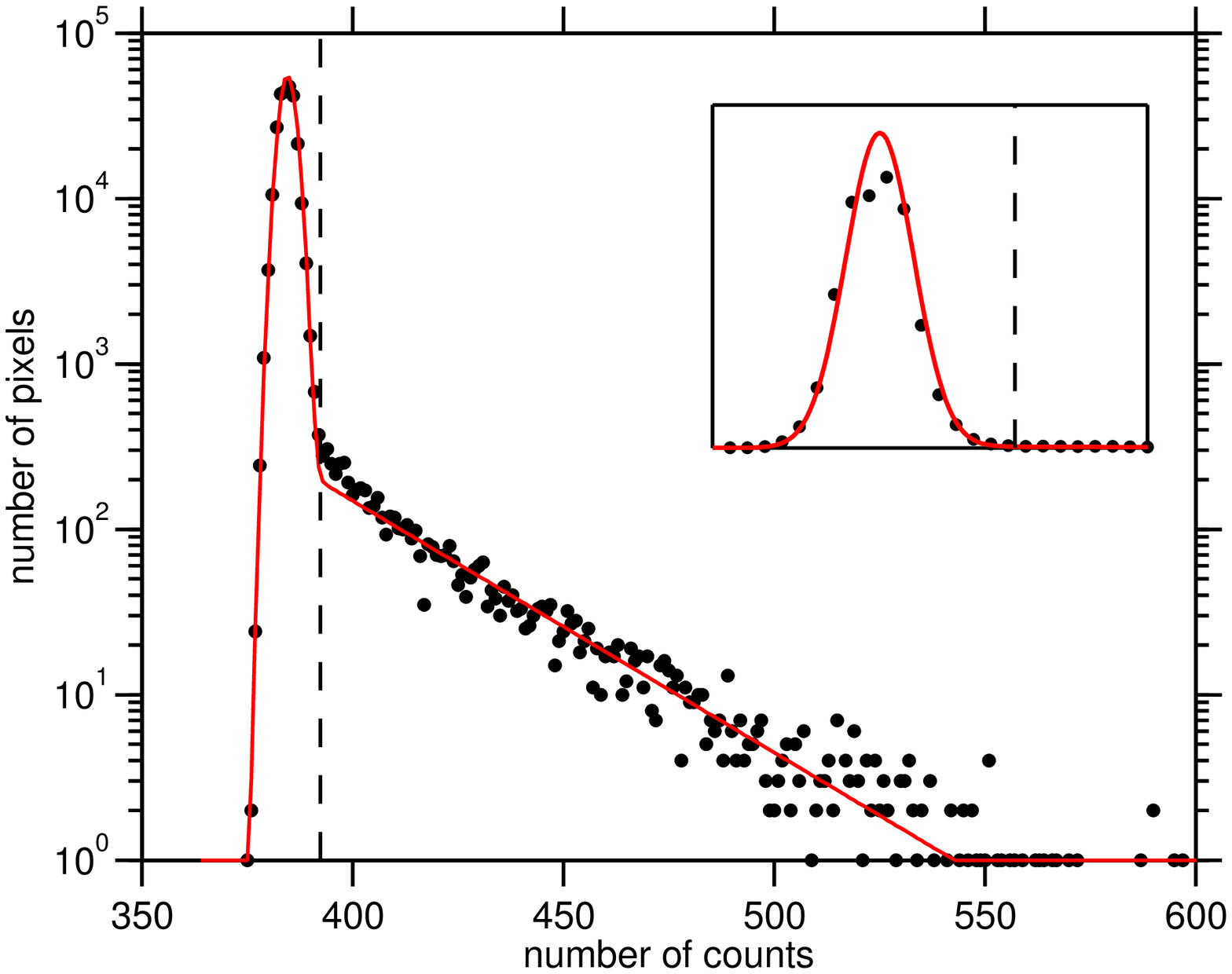}
	\caption{Distribution of the number of counts per pixel in a single image acquired without any light ($p_1=1-p_0\simeq0.02$). The detected signal is only due to clock-induced charges. The plot is in semilog scale. The red line is a fit of the distribution (see text). The dotted line corresponds to the threshold of the readout cut off. The inset shows the same distribution in linear scale.}
	\label{fig:emccd_calib}
\end{figure}

In order to retrieve an accurate estimate of $n_p$ out of the signal $S_0$, a careful calibration of the three parameters $g$, $\sigma_{ro}$ and $b$ is necessary. To obtain these values, one can fit the distribution of the number of counts per pixel on a single image. This distribution can generally be approximated by

\begin{equation}\label{eq:dist_sig}
\mathcal{S}\left(S_0;b,\sigma_{ro},g,p_i\right) =\mathcal{N}(b,\sigma_{ro}^2)\ast\left(p_0\delta\left( x\right)+\sum_{i>0} p_i \mathcal{D}\left(n_s;i,g\right) \right),
\end{equation}
where $p_i$ represents the probability to have $n_p=i$ primary electrons in a pixel, $N(b,\sigma_{ro}^2)$ is a normal distribution of mean $b$ and variance $\sigma_{ro}^2$ which accounts for the effect of the readout noise and  $\ast$ is the convolution product. In the case of an image with very low signal level ($p_0\gg p_1\gg p_i$, $i>1$), one can neglect every $p_i$ for $i>1$ in equation~\ref{eq:dist_sig}. $\mathcal{S}$ can then be calculated analytically and in turn be used to fit experimental distributions. A typical result is shown in figure~\ref{fig:emccd_calib}.

As this fitting technique can not generally be applied with our typical signal level, as light scattered by the atoms is present, each experimental measurement involves the acquisition of two successive images, one of these being used as a calibration image and hence acquired without any light. The other image can in turn be treated using the fitted value of the parameters $b$, $\sigma_{ro}$ and $g$. We first subtract the baseline $b$ to each pixel signal $S_0$. To get rid of the readout noise, we then apply a certain threshold, typically $4\sigma_{ro}$, to the remaining signal. We finally divide by the measured gain $g$. These three operations give us on average the correct initial number of photo-electrons in each pixel.

\section*{References}

\bibliographystyle{prsty}

\begin{thebibliography}{10}
\providecommand{\eprint}[2][]{\emph{Preprint} #2}

\bibitem{HanburyBrown1956}
Hanbury~Brown R and Twiss R~Q {\em Nature} 1956 {\bf 177}  27

\bibitem{Bloch2008}
Bloch I, Dalibard J, and Zwerger W {\em Reviews of Modern Physics} 2008 {\bf
  80}  885 , and references therein

\bibitem{Oettl2005}
\"Ottl A, Ritter S, K\"ohl M, and Esslinger T {\em Phys. Rev. Lett.} 2005 {\bf
  95}  090404

\bibitem{Schellekens2005}
Schellekens M, Hoppeler R, Perrin A, Gomes J~V, Boiron D, Aspect A, and
  Westbrook C~I {\em Science} 2005 {\bf 310}  638

\bibitem{Schlosser2001}
Schlosser N, Reymond G, Protsenko I, and Grangier P {\em Nature} 2001 {\bf 411}
   1024

\bibitem{Kuhr2001}
Kuhr S, Alt W, Schrader D, Muller M, Gomer V, and Meschede D {\em Science} 2001
  {\bf 293}  278

\bibitem{Nelson2007}
Nelson K~D, Li X, and Weiss D~S {\em Nat Phys} 2007 {\bf 3}  556

\bibitem{Bakr2009}
Bakr W~S, Gillen J~I, Peng A, Foelling S, and Greiner M  2009
  \eprint{0908.0174}

\bibitem{Bondo2006}
Bondo T, Hennrich M, Legero T, Rempe G, and Kuhn A {\em Optics Communications}
  2006 {\bf 264}  271

\bibitem{Wilzbach2009}
Wilzbach M, Heine D, Groth S, Liu X, Raub T, Hessmo B, and Schmiedmayer J {\em
  Opt. Lett.} 2009 {\bf 34}  259

\bibitem{Gericke2008}
Gericke T, Wurtz P, Reitz D, Langen T, and Ott H {\em Nature Physics} 2008 {\bf
  4}  949

\bibitem{Lett1988}
Lett P~D, Watts R~N, Westbrook C~I, Phillips W~D, Gould P~L, and Metcalf H~J
  {\em Phys. Rev. Lett.} 1988 {\bf 61}  169

\bibitem{Esslinger1996}
Esslinger T, Sander F, Weidemüller M, Hemmerich A, and Hänsch T~W {\em Phys.
  Rev. Lett.} 1996 {\bf 76}  2432

\bibitem{Folman2002}
Folman R, Kr{\"u}ger P, Schmiedmayer J, Denschlag J, and Henkel C {\em Adv. At.
  Mol. Opt. Phys.} 2002 {\bf 48}  263

\bibitem{Trinker2008a}
Trinker M, Groth S, Haslinger S, Manz S, Betz T, Schneider S, Bar-Joseph I,
  Schumm T, and Schmiedmayer J {\em Appl. Phys. Lett.} 2008 {\bf 92}  254102

\bibitem{Hofferberth2008}
Hofferberth S, Lesanovsky I, Schumm T, Imambekov A, Gritsev V, Demler E, and
  Schmiedmayer J {\em Nature Physics} 2008 {\bf 4}  489

\bibitem{Schumm2005b}
Schumm T, Hofferberth S, Andersson L~M, Wildermuth S, Groth S, Bar-Joseph I,
  Schmiedmayer J, and Kr{\"u}ger P {\em Nature Phys.} 2005 {\bf 1}  57

\bibitem{Wildermuth2004}
Wildermuth S, Kr{\"u}ger P, Becker C, Brajdic M, Haupt S, Kasper A, Folman R,
  and Schmiedmayer J {\em Phys. Rev. A} 2004 {\bf 69}  030901(R)

\bibitem{Ketterle1998}
Ketterle W, Durfee D, and Stamper-Kurn D  {1999} in {\em {Bose-Einstein
  condensation in atomic gases}} Vol.~{140} of {\em {Proceedings of the
  international shool of physics Enrico Fermi}} edited by {Inguscio, M and
  Stringari, S and Wieman, CE} ({Amsterdam, Netherlands}: {IOS Press}), p.\
  {67}, and references therein

\bibitem{Alt2002}
Alt W {\em Optik} 2002 {\bf 113}  142

\bibitem{Basden2003}
Basden A, Haniff C, and Mackay C {\em Monthly Notices of the Royal Astronomical
  Society} 2003 {\bf 345}  985

\bibitem{Robbins2003}
Robbins M and Hadwen B {\em IEEE T. Electron. Dev.} 2003 {\bf 50}  1227

\bibitem{Steck2001}
Steck D~A Rubidium 87 D Line Data http://steck.us/alkalidata 2001

\bibitem{Hadar1997}
Hadar O, Dogariu A, and Boreman G~D {\em Appl. Opt.} 1997 {\bf 36}  7210

\bibitem{Dettmer2001}
Dettmer S {\it et~al.} {\em Phys. Rev. Lett.} 2001 {\bf 87}  160406

\bibitem{Imambekov2009a}
Imambekov A, Mazets I~E, Petrov D~S, Gritsev V, Manz S, Hofferberth S, Schumm
  T, Demler E, and Schmiedmayer J  {\em Phys. Rev. A} 2009 {\bf 80}  033604

\bibitem{Altman2004}
Altman E, Demler E, and Lukin M~D {\em Phys. Rev. A} 2004 {\bf 70}  013603

\bibitem{Gerbier2004a}
Gerbier F, Thywissen J~H, Richard S, Hugbart M, Bouyer P, and Aspect A {\em
  Phys. Rev. A} 2004 {\bf 70}  013607

\bibitem{Hofferberth2007b}
Hofferberth S, Lesanovsky I, Fischer B, Schumm T, and Schmiedmayer J {\em
  Nature} 2007 {\bf 449}  324

\bibitem{Petrov2000}
Petrov D~S, Shlyapnikov G~V, and Walraven J~T~M {\em Phys. Rev. Lett.} 2000
  {\bf 85}  3745

\bibitem{Mazets2008}
Mazets I~E, Schumm T, and Schmiedmayer J {\em Phys. Rev. Lett.} 2008 {\bf 100}
  210403

\bibitem{Jeltes2007}
Jeltes T {\it et~al.} {\em Nature} 2007 {\bf 445}  402

\bibitem{Perrin2007}
Perrin A, Chang H, Krachmalnicoff V, Schellekens M, Boiron D, Aspect A, and
  Westbrook C~I {\em Phys. Rev. Lett.} 2007 {\bf 99}  150405

\bibitem{Greiner2005}
Greiner M, Regal C~A, Stewart J~T, and Jin D~S {\em Phys. Rev. Lett.} 2005 {\bf
  94}  110401

\bibitem{Hellweg2003}
Hellweg D, Cacciapuoti L, Kottke M, Schulte T, Sengstock K, Ertmer W, and Arlt
  J~J {\em Phys. Rev. Lett.} 2003 {\bf 91}  010406

\bibitem{Rom2006}
Rom T, Best T, van Oosten D, Schneider U, F\"olling S, Paredes B, and Bloch I
  {\em Nature} 2006 {\bf 444}  733

\bibitem{Andrews1997a}
Andrews M~R, Kurn D~M, Miesner H~J, Durfee D~S, Townsend C~G, Inouye S, and
  Ketterle W {\em Phys. Rev. Lett.} 1997 {\bf 79}  553

\bibitem{Reinaudi2007}
Reinaudi G, Lahaye T, Wang Z, and Guéry-Odelin D {\em Opt. Lett.} 2007 {\bf 32}
   3143

\bibitem{Andrews1997}
Andrews M~R, Townsend C~G, Miesner H~J, Durfee D~S, Kurn D~M, and Ketterle W
  {\em Science} 1997 {\bf 275}  637

\bibitem{Gu2000}
Gu M  1999{\em Advanced Optical Imaging Theory} ({New York}: Springer)

\end{thebibliography}

\end{document}